\documentstyle[12pt,psfig,epsf]{article}
\newcommand{\be}{\begin{equation}}
\newcommand{\ee}{\end{equation}}
\newcommand{\prr}{\prime}

\newcommand{\f}{\frac}

\newcommand{\bt}{\beta}

\def\hybrid{\topmargin -20pt    \oddsidemargin 0pt
        \headheight 0pt \headsep 0pt
        \textwidth 6.25in       % A4 paper
        \textheight 9.5in       % A4 paper
        \marginparwidth .875in
        \parskip 5pt plus 1pt   \jot = 1.5ex}

\hybrid

\def\baselinestretch{1.2}

\catcode`\@=11

\def\marginnote#1{}
\newcount\hour
\newcount\minute
\newtoks\amorpm
\hour=\time\divide\hour by60
\minute=\time{\multiply\hour by60 \global\advance\minute by-\hour}
\edef\standardtime{{\ifnum\hour<12 \global\amorpm={am}%
        \else\global\amorpm={pm}\advance\hour by-12 \fi
        \ifnum\hour=0 \hour=12 \fi
        \number\hour:\ifnum\minute<10 0\fi\number\minute\the\amorpm}}
\edef\militarytime{\number\hour:\ifnum\minute<10 0\fi\number\minute}
\def\draftlabel#1{{\@bsphack\if@filesw {\let\thepage\relax
   \xdef\@gtempa{\write\@auxout{\string
      \newlabel{#1}{{\@currentlabel}{\thepage}}}}}\@gtempa
   \if@nobreak \ifvmode\nobreak\fi\fi\fi\@esphack}
        \gdef\@eqnlabel{#1}}
\def\@eqnlabel{}
\def\@vacuum{}
\def\draftmarginnote#1{\marginpar{\raggedright\scriptsize\tt#1}}

\def\draft{\oddsidemargin -.5truein
        \def\@oddfoot{\sl preliminary draft \hfil
        \rm\thepage\hfil\sl\today\quad\militarytime}
        \let\@evenfoot\@oddfoot \overfullrule 3pt
        \let\label=\draftlabel
        \let\marginnote=\draftmarginnote
   \def\@eqnnum{(\theequation)\rlap{\kern\marginparsep\tt\@eqnlabel}%
\global\let\@eqnlabel\@vacuum}  }

\def\preprint{\twocolumn\sloppy\flushbottom\parindent 2em
        \leftmargini 2em\leftmarginv .5em\leftmarginvi .5em
        \oddsidemargin -.5in    \evensidemargin -.5in
        \columnsep .4in \footheight 0pt
        \textwidth 10.in        \topmargin  -.4in
        \headheight 12pt \topskip .4in
        \textheight 6.9in \footskip 0pt
        \def\@oddhead{\thepage\hfil\addtocounter{page}{1}\thepage}
        \let\@evenhead\@oddhead \def\@oddfoot{} \def\@evenfoot{} }

\def\numberbysection{\@addtoreset{equation}{section}
        \def\theequation{\thesection.\arabic{equation}}}

\def\underline#1{\relax\ifmmode\@@underline#1\else
        $\@@underline{\hbox{#1}}$\relax\fi}

\def\titlepage{\@restonecolfalse\if@twocolumn\@restonecoltrue\onecolumn
     \else \newpage \fi \thispagestyle{empty}\c@page\z@
        \def\thefootnote{\fnsymbol{footnote}} }

\def\endtitlepage{\if@restonecol\twocolumn \else \newpage \fi
        \def\thefootnote{\arabic{footnote}}
        \setcounter{footnote}{0}}  %\c@footnote\z@ }

\catcode`@=12
\relax

\def\figcap{\section*{Figure Captions\markboth
        {FIGURECAPTIONS}{FIGURECAPTIONS}}\list
        {Figure \arabic{enumi}:\hfill}{\settowidth\labelwidth{Figure
999:}
        \leftmargin\labelwidth
        \advance\leftmargin\labelsep\usecounter{enumi}}}
 \relax
\def\tablecap{\section*{Table Captions\markboth
        {TABLECAPTIONS}{TABLECAPTIONS}}\list
        {Table \arabic{enumi}:\hfill}{\settowidth\labelwidth{Table
999:}
        \leftmargin\labelwidth
        \advance\leftmargin\labelsep\usecounter{enumi}}}
 \relax
\def\reflist{\section*{References\markboth
        {REFLIST}{REFLIST}}\list
        {[\arabic{enumi}]\hfill}{\settowidth\labelwidth{[999]}
        \leftmargin\labelwidth
        \advance\leftmargin\labelsep\usecounter{enumi}}}
 \relax
\makeatletter
\newcounter{pubctr}
\def\publist{\@ifnextchar[{\@publist}{\@@publist}}
\def\@publist[#1]{\list
        {[\arabic{pubctr}]\hfill}{\settowidth\labelwidth{[999]}
        \leftmargin\labelwidth
        \advance\leftmargin\labelsep
        \@nmbrlisttrue\def\@listctr{pubctr}
        \setcounter{pubctr}{#1}\addtocounter{pubctr}{-1}}}
\def\@@publist{\list
        {[\arabic{pubctr}]\hfill}{\settowidth\labelwidth{[999]}
        \leftmargin\labelwidth
        \advance\leftmargin\labelsep
        \@nmbrlisttrue\def\@listctr{pubctr}}}
 \relax
\makeatother
\newskip\humongous \humongous=0pt plus 1000pt minus 1000pt

\newif\ifdtup

\relax

\def\be{\begin{equation}}
\def\ee{\end{equation}}
\def\ba{\begin{eqnarray}}
\def\ea{\end{eqnarray}}

\def\b{\beta}

\def\L{\Lambda}

\def\no{\noindent}

\def\IR{\relax{\rm I\kern-.18em R}}

\def\IR{\relax{\rm I\kern-.18em R}}
\def\inv{^{\raise.15ex\hbox{${\scriptscriptstyle -}$}\kern-.05em 1}}

\def\tL{{\tilde L}}

\begin{document}
%\draft

\renewcommand{\theequation}{\arabic{equation}}

\newcommand{\beq}{\begin{equation}}
\newcommand{\eeq}[1]{\label{#1}\end{equation}}
\newcommand{\ber}{\begin{eqnarray}}
\newcommand{\eer}[1]{\label{#1}\end{eqnarray}}
\newcommand{\eqn}[1]{(\ref{#1})}
\begin{titlepage}
\begin{center}

%Phys. Lett. {\bf B432} (1998) 365
\hfill NTUA-95/00\\
\hfill hep--th/0007079
\\

\vskip 1.2in

{\large \bf Lattice Evidence for Gauge Field Localization on a Brane}

\vskip 0.6in

{\bf P. Dimopoulos, K. Farakos, A. Kehagias and G. Koutsoumbas}
%\phantom{x}  and\phantom{x} {}
\vskip 0.1in
{\em Physics Dept., National Technical Univ.,  \\
     Zografou Campus, 157 80 Athens, Greece}\\
%{\tt kehagias@mail.cern.ch}}\\
\vskip .2in

\end{center}

\vskip .6in

\centerline{\bf Abstract }

\no
We examine the problem of gauge-field localization in higher-dimensional 
gauge theories. In particular, we study a five-dimensional $U(1)$ by lattice techniques 
and we find that gauge fields can 
indeed be localized. 
Two models are considered. The first one has 
anisotropic couplings 
independent of each other and of the coordinates.
It can be realized on a homogeneous but anisotropic flat Euclidean space.
The second model has couplings depending on the extra transverse fifth 
direction. This model can be realized by a $U(1)$ gauge  theory on the 
Randall-Sundrum background. We find that in both models
a new phase exists, the layer phase, in which a massless photon is localized 
on the four-dimensional layers. We find that this phase is separated 
from the strong coupling phase by a second order phase transition. 

\vskip 0,2cm
\no

\vskip 4cm
\noindent
%CERN-TH/99-192\\
July 2000\\
\end{titlepage}
\vfill
\eject

\def\baselinestretch{1.2}
\baselineskip 16 pt
\noindent

%%%%%%%%%%%%%%%%%%%%%A generalization of target space%%%%%%%%%
\def\tT{{\tilde T}}
\def\tg{{\tilde g}}
\def\tL{{\tilde L}}

%%%%%%%%%%%%%%%%%%%

\section{Introduction}

The idea that we  live in a higher-dimensional 
space-time is not new. 
The general setting is a $(4+n)$-dimensional space-time with $n$ dimensions 
compactified and the vacuum is of the form $M^{1,3}\!\times\! X^n$. 
$M^{1,3}$ is the four-dimensional Minkowski space-time we experience and 
$X^n$ is an internal compact space. 
The compactness of the internal space is crucial since the four-dimensional
Planck mass $M_{Pl}$ is related to the $4+n$ dimensional Planck mass 
$M_{(4+n)}$ by $M_{Pl}^2=M_{(4+n)}^{2+n} V_n$  
where $V_n$ is the volume of $X^n$.
It is exactly this relation which allows for a large volume internal space
\cite{IA},\cite{ADD}. 
Thus, as we see,
 four-dimensional gravity exists as long as the volume of the internal 
space is finite. The latter requirement together with the smoothness of the 
internal space can be satisfied with a compact space $X^n$. Non-compact 
internal spaces have also been considered in the past at the cost of 
giving up smoothness \cite{GZ},\cite{Nww},\cite{CKK},\cite{KRR},\cite{WWW}. 
Indeed,  all such cases are suffering from naked 
singularities.  However, there are also cases where such singularities are
 not that bad as for example delta-function singularities to which a
 physical interpretation may be given as domain walls, strings, etc. This is 
the case  in the Randall-Sundrum (RS) model, where the 
singularities may be interpreted as a four-dimensional  domain wall, a
three-brane, embedded in a five dimensional bulk \cite{RS}.     
In that case, although the internal space may be non-compact, a 
four-dimensional graviton exists and it is localized at the three-brane.
The question one would like to answer is if in this case,
localization on the 3-brane exists for other 
fields, like gauge fields, fermions and scalars.

It is known that solitons like domain walls or strings,
 may support massless fields. In theories
with domain walls formed by a scalar field (vortex field)
for example, scalars as well as fermions with appropriate
interactions in the bulk give rise to massless scalars and chiral fermions
on the wall \cite{L}--\cite{J}. Fermions in five-dimensions in particular,
with Yukawa couplings to the vortex field, deposit a single chiral fermionic
zero mode on the four-dimensional wall \cite{CH}. Thus, eventually,
by choosing appropriate couplings of the bulk fermions to the vortex field,
it may be possible to get the Standard Model fermionic spectrum as the
zero-mode sector localized on the four-dimensional wall \cite{RuSa}.
Although scalars and fermions can easily be localized, there is no fully
satisfactory localization
mechanism for gauge fields. The reason is the following. Consider a $U(1)$
theory in five dimensions which couples to the vortex field.
The latter forms the domain wall and also breaks the $U(1)$ theory by
developing a vacuum expectation value. In addition, it is
possible that the vortex field vanish at the position of the wall.
(The particular way this is achieved is not important for the argument.)
As a result, the $U(1)$ theory is broken everywhere 
except at the position of
the wall leading to a massless four-dimensional photon localized on
 the wall. However, this localized  photon,
unfortunately, does not remain massless.
The reason is that, due to the condensation of the vortex field,
a superconducting medium is produced everywhere except at the position of the
domain wall. It is clear then that no long-range electric fields can be
produced along the wall as a result of the Meissner effect. Instead, the
magnetic flux is confined in the wall and it is spread according to the
Coulomb law. The proposals made so far for the localization of
 gauge fields, in fact reverse the above situation
\cite{BK},\cite{DS},\cite{ADD},\cite{TR}.
A dual version of it is employed by replacing the superconductor by a
confining medium with monopole condensation in the bulk \cite{giacomo}. 
In this case, electric and
magnetic fields are interchanged leading to  long-range electric fields
spreading according to Coulomb law 
within the 4-D wall (3-brane), while magnetic fields die-off exponentially
with the distance along the wall.

The RS background is a vortex in a certain sense where the 3-branes play the
role of the vortex field. In such a case, we would like to know if
there exist localized fields on the 3-brane. At first,
the appearance of a localized four-dimensional field in the RS geometry
 may be seen as follows.
The metric of the RS background, although it is 
continuous, it has discontinuous first
derivatives. This leads to a delta-function singularity in the Riemann tensor
and, consequently, to the Einstein equations. Expanding the action around
the RS background, an attractive delta-function potential
in a Schr\"odinger-like equation is produced
for the transverse dependence of the four-dimensional graviton excitation.
This delta-function attractive potential supports a unique bound state
which appears as the massless four-dimensional graviton. On top of the
massless graviton, there exists the continuum spectrum of
Kaluza-Klein  states. These states produce an extremely suppressed
contribution to the effective four-dimensional theory 
so that the usual Newton
law is produced \cite{RS}.

Similar considerations also apply in the case of bulk scalars \cite{GW},
fermions \cite{Rizzo}, \cite{GP} and gauge fields \cite{Rizzo},\cite{P}.
 In searching for localized four-dimensional fields, the
 equations of motion of the bulk fields which are coupled to the
background geometry are solved . 
These equations have solutions which represent
localized four-dimensional massless and/or massive fields for scalars and
fermions. However, there are no localized solutions for gauge fields.
This negative result does not exclude the possibility of gauge-field
localization. It rather indicates 
 that if localized gauge fields 
exist, the gauge theory should not be in a Coulomb phase in the bulk.
Intuitively, one expects that a confining five-dimensional theory may give
rise to a four-dimensional gauge theory in the Coulomb phase.
We will examine exactly this possibility by lattice methods, since there
are no other techniques available. In particular, we will consider a $U(1)$
gauge theory in the RS background and we will show that massless photons
 exist in four dimensions while there is confinement in the 
transverse fifth direction. The reason for such behaviour is that the 
effect of the 
RS-background or a general anti-de Sitter $(AdS_5)$ background on the 
$U(1)$ gauge theory is to provide the 
gauge theory with  a different gauge coupling in the fifth direction. 
In this case, we know, by the work of Fu and Nielsen 
\cite{NF} (see also \cite{Altes}) 
that a new phase of the five-dimensional gauge theory 
may exists. This phase, the layered phase, is confining in the fifth 
direction and Coulomb 
within the four-dimensional layers. We claim that this layered phase is 
responsible for the localization of the gauge fields in four dimensions. 
Note that the layered phase for non-abelian gauge theories also 
exists in six dimensions \cite{BR}. There is a conceptual problem
with the layered phase that shows up in \cite{BR}. Within the layers 
we have a Coulomb phase, while the strong coupling characterizes the
coupling of the layers. Thus, insofar as localization is concerned, this
layered phase is fine, but the space  which is supposed to become the 
``physical" space in which we live is Coulombic, that is it does not
have confinement. Thus, unfortunately, the non-Abelian pure gauge theories 
do not have an interesting layered phase. A possible way out could be 
to introduce scalar matter fields in the non-Abelian theory \cite{SFDKK}. 
In this
case it is conceivable that things could be arranged so that one may 
have layers in the Higgs phase (which represents the world we are living
in) coupled strongly with one another, thus localizing gauge fields on
the branes.

%The reason for choosing a $U(1)$
%theory is not only for its relative simplicity but
%also because a non-abelian gauge theory in five dimensions does not have a
%confining phase.

\section{The geometry of the AdS space and the RS setup}

We will recall here some well-known results for the
 five-dimensional
$AdS_5$ space-time. (Generalization to other dimensions is
 straightforward.) $AdS_5$ is a five-dimensional maximally symmetric
space-time with negative cosmological constant $\L<0$. It
 can be viewed as the hyperboloid
\be
-X_{-1}^2-X_0^2+X_1^2+\ldots+X_4^2=R^2 \, , \label{emb}
\ee
of radius R in a six-dimensional flat space of $(-1,-1,1,1,1,1)$ signature.
 Topologically, the space is $S^1\times
{\bf R}^{4}$ where the $S^1$ represents closed time-like curves. However, by
unwraping the $S^1$ we obtain a space -time with no closed time-like curves
and this is the space which we will normally call $AdS_5$. The induced
metric on the hyperboloid, in appropriate coordinates, takes the form
\be
ds^2=e^{-2 r/R}\left(-dt^2+dx_1^2+dx_2^2+dx_3^2\right)
+dr^2\, ,~0 \le r < \infty.
\label{met1}
\ee
 In horospheric coordinates ($t,x_1,x_2,x_3,
u=e^{-r/R})$, the AdS metric takes the form
\be
ds^2=R^2u^2\left(-dt^2+dx_1^2+dx_2^2+dx_3^2\right)+R^2{du^2\over u^2}\, ,
\label{met2}
\ee
which is the form appearing as the near-horizon limit of D3-branes.

The Riemann tensor of the $AdS_5$ space is
\be
R_{MNPQ}=-{1\over R^2}
\left(g_{MP}g_{NQ}-g_{MQ}g_{NP}\right)\, , ~~~~M,N,...=0,...4,
\, 
\ee
shows clearly that $AdS_5$ is a maximally symmetric space-time with 
negative cosmological constant $\L$ proportional to $-1/R^2$.
Its isometry group is $SO(2,4)$  as can be seen from the embedding
(\ref{emb}).
Its boundary is  the so-called compactified
Minkowski space-time $\tilde{M}^{1,3}$. It is the Minkowski
space-time  $M^{1,3}$ at $r\to -\infty$ plus the point at $r\to \infty$.
The isometry group  $SO(2,4)$
 acts as the conformal group on $\tilde{M}^{1,3}$
and it is clear from the form of the metric (\ref{met1}) (or (\ref{met2})),
 that there is four-dimensional
Poincar\'e symmetry. This is the
group-theoretic reason for the AdS/CFT correspondence \cite{Mald},
namely, the twofold
interpretation of the $SO(2,n)$ as a symmetry group of $AdS_{n+1}$ or
as the conformal group of its boundary  $\tilde{M}^{1,n-1}$.
According to the AdS/CFT correspondence, the type IIB supergravity theory on
$AdS_5\times S^5$ is dual to ${\cal N}=4$ SUSY $SU(N)$
YM theory at large $N$ in 4-D Minkowski space. 
The YM theory lives in the boundary of
$AdS_5$ and there is a precise correspondence of bulk and boundary terms
\cite{Mald},\cite{KP},\cite{WW}.
In this setup, gravity lives in the bulk of $AdS_5$ while the gauge theory
lives in the boundary $M^{1,3}$.
%This is due to the fact, that the gauge sector comes
%from the open strings which end on the D3-brane while the graviton originates
%from closed-strings living in the bulk.

``Boundary" gravitons may exist by appropriate modification of the metric
(\ref{met1}). This is the programme originated in \cite{RS} and further
elaborated by many others. The metric which allows for a localized
``boundary" graviton is
\be
ds^2=e^{-2k r_c|\phi|}\left(-dt^2+dx_1^2+dx_2^2+dx_3^2\right)
+r_c^2d\phi^2\, .
\label{RSm}
\ee
In the above relation $r_c$ is a compactification scale and $k$ is
related to the bulk cosmological constant.
In the original model, the coordinate $\phi$
parametrizes $S^1/Z_2$ so that $-\pi\leq \phi\leq\pi$ with $\phi$ identified
with $-\phi$.
The embedding of the Randall-Sundrum in supergravity was given
in \cite{K}.
 Calculating the energy momentum tensor for the Randall-Sundrum metric
(\ref{RSm}) we find that there exist two domain walls (3-branes)
at $\phi=0, \pi$ with positive and negative
tension, respectively.
The four-dimensional Planck mass $M_{Pl}$ is then
\be
M_{Pl}^2= {M^3\over k}(1-e^{-2kr_c\pi}) \label{mm}
\ee
where $M$ is the Planck mass in five dimensions.
In addition, a massive field with mass parameter $m_0$
confined in the 3-brane at $\phi=\pi$, appears to have a physical mass
$m=m_0e^{-kr_c\pi}$. In view of this, weak scale is generated from a
fundamental Planck scale if $kr_c\sim 12$ providing an ``exponential"
solution to the hierarchy problem \cite{RS}.

 The unphysical negative-tension brane now can be 
safely moved to infinity by sending the ``compactification" radius
$r_c\to \infty$ leaving a semi-infinite space.
  It is  obvious from eq.(\ref{mm}),
 that the limit $r_c\to \infty $ can be taken which still
results in a finite four-dimensional Planck mass. In this case, the
negative-tension brane is moved to infinity and  exponential
hierarchy is lost.
A finite $r_c$ clearly leads to a four-dimensional dynamics with
all fields  accompanying by their KK states. However,
by sending $r_c$ to infinity, the existence of four-dimensional dynamics is
not obvious \cite{mavleo}. 
Contrary, all fields are expected to live in the bulk of
space-time with no localized zero modes on the 3-brane. If this was the case,
the whole construction would be of no particular interest. It happens 
however,
that bulk fields have well localized zero modes on the brane in a novel way.
It should be noted however here, that non-compact compactifications, i.e.,
``compactifications" where the internal space is a non-compact rather a
compact space appears before as well but in most of the cases they were
accompanied by naked singularities \cite{GZ},\cite{Nww},\cite{CKK},
\cite{KRR}.
The interpretation of the latter, made
their existence quite obscure \cite{WWW}.
%All results seems to confirm
%that bulk scalars and fermions can be localized while bulk gauge fields can
%not.
%This is similar to what happens in the
%vortex case. For the domain wall we discussed above, a

\section{5D $U(1)$ Gauge Fields and their Lattice Action}

Following the discussion of the previous section 
let us consider an abelian scalar model with fermions in a
five-dimensional $AdS_5$ background. The action of such a theory
is given by
\be
S=S_{gauge}+S_{scalar}+S_{fermion}\, ,
\ee
where
\ba
&&S_{gauge}=-{1\over
4 g_5^2}\int d^5 x \sqrt{-g}F_{MN} F_{KQ}g^{MK}g^{NQ}\nonumber \\
&&=\int d^4x dr
\left(-{1\over
4g_5^2}F_{\mu\nu}F_{\kappa\lambda}\eta^{\mu\kappa}\eta^{\nu\lambda}-{1\over
2g_5^2}e^{-2 r/R}F_{\mu 5}F_{\nu 5}\eta^{\mu\nu}\right)\, ,
\label{gauge}
\ea
is the action for the $U(1)$ gauge field where $g_5$ is the
dimensionful gauge coupling in five dimensions.
\ba
&&S_{scalar}=\int d^5 x \sqrt{-g}\left(-D_M\Phi
D_N\Phi^*g^{MN}-V(\Phi)\right)\nonumber \\
&&=\int d^4 x dr \left(e^{-2 r/R}D_\mu\Phi D_\nu \Phi^* \eta^{\mu\nu}+e^{-4r/R}
D_5\Phi D_5\Phi^*-e^{-4r/R}V(\Phi)\right)\, ,
\ea
is the action for the scalar field with potential term and
$D_M=\partial_M+i A_M$. Finally, the fermion action is
\be
S_{fermion}=\int d^5x \sqrt{-g}\left(
\bar{\Psi}\Gamma^M\hat{D}_M\Psi\right)
\ee
where $\hat{D}_M=D_M+{1\over 4}\omega_{MAB}\Gamma^{AB}$ is the
spin-covariant derivative.

We will study below by lattice techniques only the gauge theory without matter fields.
A similar system with a scalar field included is under
consideration \cite{SFDKK}.
It is a trivial matter to analytically continue the Minkowski space action
(\ref{gauge}) to Euclidean space so that we get
\ba
&&S_{gauge}^{E}=\int d^5x
\left({1\over
4g_5^2}F_{\mu\nu}F_{\mu\nu}+{1\over
2g_5^2}e^{-2 |x_T|/R}F_{\mu T}F_{\mu T}\right)\,
,~~~\mu,~\nu=1,...,4\, .
\label{gauge1}
\ea
We have defined the original coordinate $r$ with $0\leq r<\infty$ as $r=x_T$
 (T denotes the fifth-transverse direction), where now $-\infty<x_T<\infty$,
and we have set the absolute value in the exponent. In this way,
we have essentially a gauge theory on a RS background. We claim
that, this form of the action captures all the essential features
of a gauge theory on this background and its study will reveal
non-perturbative dynamics and possible different phases of this
model. In particular, we expect that the localization of the gauge
theory on a four-dimensional continuum will possibly be answered
in such a setup.
It is obvious from the action (\ref{gauge1}) that the gauge
coupling in the fifth direction is bigger (depending on $x_T$)
from the coupling in the four-dimensional space-time.

Let us remind the reader of a couple of basic facts about the 
lattice formulation of a gauge field theory.
The link variables take the form:
\be
U_M(x)=\{U_\mu=e^{ia_s {\overline A}_\mu},~U_T(x)=e^{ia_T {\overline A}_T} \}
\ee
while the plaquette variables are
$$
U_{\mu\nu}(x)=U_\mu(x)U_\nu(x+a_s {\hat \mu})
{U_\mu}^\dagger(x+a_s {\hat \nu})
{U_\nu}^\dagger(x)
$$
$$
U_{\mu T}(x)=U_\mu(x)U_T(x+a_s{\hat \mu}){U_\mu}^\dagger(x+a_T {\hat T})
{U_T}^\dagger(x)\, ,
$$
where $a_s,a_T$ are the lattice spacings in the 4-D space and the
transverse direction, respectively and the quantities ${\overline A}_\mu$ and
${\overline A}_T$ represent the corresponding gauge potentials.

We are now going to write down a discretized version of the gauge action
on a cubic five-dimensional lattice permitting
different gauge couplings $\b, \b'$ between the 4-D space and the
transverse fifth direction. We will treat two different models in this 
work, so we describe them separately.

{\bf Model I:}

The Wilson action for pure $U(1)$ in five dimensions with
anisotropic couplings takes the form:
\be
S_{gauge}^I=\b \sum_{x,1\leq \mu<\nu\leq 4}(1-Re\ U_{\mu\nu}(x))+
\b' \sum_{x,1\leq \mu\leq 4 }(1-Re\ U_{\mu T}(x))\, .
\label{lat}
\ee
In this model we will assume that the gauge couplings $\b,\b'$ 
are generically independent from each other and
from the fifth coordinate $x_T$. This guarantees that there is an unbroken
four-dimensional Poincar\'e invariance in the continuum limit as
long as $\b$ and $\b'$ are different. However, this symmetry is enhanced to a 
five-dimensional Poincar\'e invariance at the special point $\b=\b'$.

In the naive continuum limit $(a_s,a_T\to 0)$ the lattice action
for model I turns out to be:
\be
S_{gauge}^I={\bt \over 2} \sum_{x,1\leq \mu<\nu\leq 4} F_{\mu\nu}^2
+\f{\bt^\prr}{2} \sum_{ x,1\leq \mu\leq 4 } F_{\mu T}^2 + {\cal
O}(a^5)\, .
\ee
where 
$F_{\mu \nu}(x) \equiv A_\nu(x+a_s {\hat \mu})-A_\nu(x)
-A_\mu(x+a_s {\hat \nu})+A_\mu(x),$
$F_{\mu T}(x) \equiv A_T(x+a_s {\hat \mu})-A_T(x)
-A_\mu(x+a_t {\hat T})+A_\mu(x).$
The next step is to rewrite the action in terms
of the continuum fields (denoted by a bar):
$$
~~ A_\mu = a_s {\overline A_\mu}~(\mu=1,2,3,4),~~A_T = a_T {\overline A_T}.
$$
This means that the transverse-like part of the pure gauge action is
rewritten in the form: $$\f{\beta^\prr a_{T}}{2 a_s^2} \sum a_{s}^{4}a_{T}
\sum_{\mu=1,2,3,4} ({\overline F}_{\mu T})^{2}
\rightarrow \f{\beta^\prr a_{T}}{2 a_s^2} \int d^{5}x
\sum_{\mu=1,2,3,4} ({\overline F}_{\mu T})^{2}.$$
On the other hand the space--like part is:
$$\f{\beta}{2 a_T} \sum a_s^4 a_T
\sum_{1 \le \mu < \nu \le 4} ({\overline F_{\mu \nu}})^2
\rightarrow \beta \frac{a_s^2}{2 a_T} \int d^5 x
({\overline F_{\mu \nu}})^2.$$
If we define
\be
\beta \equiv \frac{a_{T}}{g_{5}^{2}},
~~\beta^\prr \equiv \frac{a_s^2}{g_{5}^{2} a_{T}},
\ee
the resulting continuum action reads:
$$\f{1}{2 g_{5}^{2}} \int d^5 x {\left [ 
\sum_{1 \le \mu < \nu \le 4}({\overline F_{\mu \nu}})^{2}
+\sum_{\mu=1,2,3,4} ({\overline F_{\mu T}})^{2} \right ]}$$

The above action takes the standard form in the continuum
\be
 S_{gauge}^I=\int d^5x {1\over 2 g_5^2}F_{MN}^2\, .
\ee

Note that $g_5^2$ has dimensions of length and is related to 
a characteristic scale for five dimensions.

This expression does not exhibit any anisotropy at all.
However the na\"{\i}vet\'e of this approach will be manifest by
the results we present below that indicate that the anisotropy
may survive in the continuum limit.

\vspace{1cm}

{\bf Model II:}

\vspace{1cm}

Let us now turn to a second model, which we will call model
II and which can be viewed as a discretization of a gauge theory in the
RS background. In this model, the gauge coupling $\b'$ depends on
the fifth coordinate $x_T=n_T a_T$ and the action is
\be
S_{gauge}^{II}=\b \sum_{x,1\leq \mu<\nu\leq 4}(1-Re\ U_{\mu\nu}(x,y))+
\b \sum_{x,1\leq \mu\leq 4 }e^{-|n_T|\lambda}(1-Re\ U_{\mu T}(x,y))\, .
\label{lat1}
\ee
where $\lambda=2 a_T/R$. The  na\"ive continuum limit of the model II is just the action
(\ref{gauge1}). Clearly here $\b,\b'$ are not independent, as in
model I, but they are related through the equation:
\be
\b'=\b e^{-|n_T|\lambda}\, .
\ee

\vspace{1cm}

{\bf Short description of the phases:}

\vspace{1cm}

A very important ingredient of the lattice treatment of both models 
is the appearance of the so called layered phase; this is a good point 
to explain some of its characteristics.
Let us suppose that we start with large values for $\bt$ and $\bt^\prr,$
so the model lies in a Coulomb phase in five dimensions. There is a
Coulomb force between two test charges in this phase. Now consider what
will happen when one keeps $\bt$ constant, but lets $\bt^\prr$ take smaller
and smaller values. Nothing will change in the four directions
that have to do with $\bt,$ so the force will still be Coulomb-like;
however, the force between the test charges in the fifth
direction will increase and will eventually become confining when
$\bt^\prr$ becomes small enough. It is well known that the potential between
heavy test charges is closely connected with the Wilson loops. According to
the above scenario, the Wilson loops are expected to behave as follows:
\newpage

%\begin{enumerate}
%%%%%%%%%%%%%%%%%%%%%%%%%%%%%%%%%%%%%%%%%%%%%%%%%%%%%%%%%%%%%%%%%%%%%%%%%
\hspace{1.34cm}1.  $W_{\mu \nu}(L_1,L_2) \approx exp(-\sigma L_1 L_2) $ (strong coupling).
\vspace{0.1cm}

\hspace{1.34cm}2.  $W_{\mu \nu}(L_1,L_2) \approx exp(-\tau (L_1+L_2)) $
(Coulomb phase, $1 \le \mu, \nu \le 5.)$

\[ \left. \begin{array}{l}
3. \hspace{0.15cm}  W_{\mu \nu}(L_1,L_2) \approx exp(-\tau (L_1+L_2)) \\
\\
4. \hspace{0.15cm} W_{\mu T}(L_1,L_2) \approx exp(-\sigma^\prr L_1 L_2) 
\end{array}
\right \}  (\mbox{layered phase}, 1 \le \mu, \nu \le 4.) \] 
%\end{enumerate}
%%%%%%%%%%%%%%%%%%%%%%%%%%%%%%%%%%%%%%%%%%%%%%%%%%%%%%%%%%%%%%%
The quantities $\sigma,~\tau,~\sigma^\prr$ are positive constants.
Let us remark here that there is no layered phase with the roles of $\bt$
and $\bt^\prr$ interchanged, since the two parameters enter in a quite different
way in the action. The layered phase is due to the simultaneous
existence of Coulomb forces in the space-like directions and confining
forces in the fifth direction. We need an initial theory having 
two distinct phases, if we are to see a layered structure. This is
why we start with a (4+1)-D abelian theory. For a non-Abelian
theory, we would need at least 5+1 dimensions.

\section{Results}

Before presenting the results we give some technical details of
the simulation.
We use a 5-hit Metropolis algorithm supplemented by an 
overrelaxation method, which is based on rewriting the action in the form:
$$
C cos(\phi+\theta_\mu)
$$
Defining 
$$P_1 \equiv C cos \phi = \sum cos \chi_s + q \sum cos \chi_T, $$
$$P_2 \equiv C sin \phi = \sum sin \chi_s + q \sum sin \chi_T, $$
where $\chi_s$ represent space-like staples, while $\chi_T$ the
transverse-like ones. The parameter $q$ is the quotient
$q \equiv \f{\bt^\prr}{\bt},$ in the case of space-like $\theta_\mu$ and 1
otherwise. The new link is given by the standard formula of the overrelaxation
method \cite{creutz,brown}, 
which in our case reads: $\theta'_\mu = -\theta_\mu-2 \phi.$
The change is always accepted.
An important technical issue has been the selection of the intervals
around the old values of the link variables in which the trial new
values will lie. It is well known that these ranges depend on the 
acceptance rates; in this model we face very frequently the situation
that the acceptance rates for the space-like directions are very different
from the ones for the transverse direction, because the coupling constants 
can be very different. Thus we have to use different ranges for the
space-like directions and the transverse direction, corresponding
to the different acceptance rates.

\subsection{Model I}
Let us describe the quantities we use to spot the phase transitions 
and their orders.
Two operators that we use heavily in this work are the following:
\be
{\rm Space-like~Plaquette:~~~} {\hat P}_s \equiv \f{1}{6 N^5}
\sum_{x,1 \le \mu < \nu \le 4} \cos F_{\mu \nu}(x)
\ee
\be
{\rm Transverse-like~Plaquette:~~~} {\hat P}_T \equiv \f{1}{4 N^5}
\sum _{x,1 \le \mu \le 4} \cos F_{\mu T}(x)
\ee
We have used the mean values $P_s$ and $P_T$ of the above quantities
in performing hysteresis
loops; in addition, using these operators we have derived new quantities,
whose behaviours characterize the orders of the various phase transitions.
The quantities used in this work are:
\begin{enumerate}
\item The mean value of the space-like plaquette: $$P_s \equiv <{\hat P}_s>.$$
\item The mean value of the transverse-like plaquette: $$P_T \equiv <{\hat P}_T>.$$
\item The distribution N(${\hat P_s}$) of ${\hat P_s}.$
\item The susceptibilities of ${\hat P_s},~{\hat P_T}:$ $$ S({\hat P_s})
\equiv V (<{\hat P_s}^2>-<{\hat P_s}>^2),~S({\hat P_T}) \equiv 
V (<{\hat P_T}^2>-<{\hat P_T}>^2).$$
\item The Binder cumulant of ${\hat P_s}:$ $$ C({\hat P_s})
\equiv 1-\frac{<{\hat P_s}^4>}{3 <{\hat P_s}^2>^2}.$$
\end{enumerate}
The symbol $<...>$ denotes the statistical average.

We start by quoting from references \cite{NF}, \cite{Altes} 
the phase diagram of 
(4+1)-dimensional U(1). This is our figure 1, which contains the curves
labeled by (I), corresponding to zeroth order mean field theory, and
the curves (II), where the one-loop corrections to the mean field result
have been taken into account. The curves have similar qualitative behaviour, 
but the Monte Carlo results which follow agree better with the one-loop
results. Notice that the part of the phase transition line between the 
strong and the Coulomb phase corresponding to $\bt^\prr>1.0$ is outside
the range of validity of the strong coupling expansion used in \cite{NF}.
We observe that 
for small values of $\bt,~\bt^\prr$ we have the strong coupling phase ({\bf S}),
where all Wilson loops obey the area law, signaling confinement in 
all five directions. In the regime of large values of  $\bt,~\bt^\prr,$
that is weak coupling, we have the so-called Coulomb phase ({\bf C}), with the 
perimeter law holding for the Wilson loops in all directions. 
The new phase showing up here is the layered phase ({\bf L}), which appears for 
large values of $\bt$ (weak coupling in 4-D ``space") and small values of
$\bt^\prr$ (strong coupling in the fifth direction). As explained previously,
the 4-D space lies in the Coulomb phase, while the model confines in the
fifth direction. As already explained by the authors of reference 
\cite{Altes}, the transition between the strong and the Coulomb phases is
quite strong (this is why it is represented by a solid line), while the 
other two transitions (strong-layered and layered-Coulomb) are much
weaker. We will study a representative point from each of the three 
transitions: these points are labeled A (at $\bt^\prr=1.0$), 
B (at $\bt^\prr=0.2$) and C (at $\bt=1.4$) in figure 1.

%%%%%%%%%%%%%%%%%%%%%%%%%%%%%%%%%%%%%%%%%%%%%%%%%%%%%%%%%%%%%%%%
\begin{figure}
\centerline{\hbox{\psfig{figure=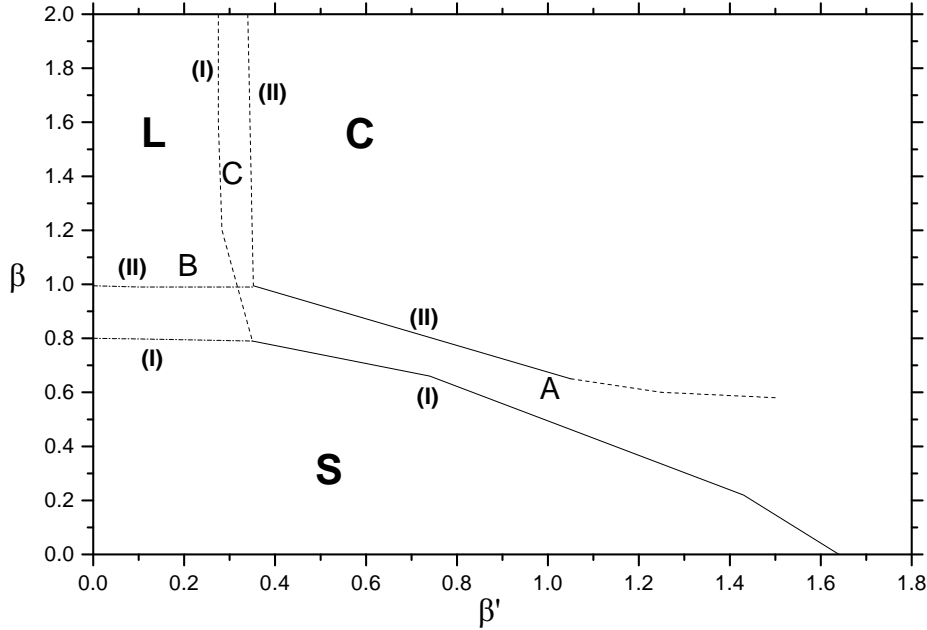,height=10cm}}}
\caption[f1]{Phase diagram of pure U(1) gauge theory in 4+1 dimensions 
with anisotropic couplings. Points A, B and C 
will be studied in detail in the sequel.}
\label{f1}
\end{figure}
%%%%%%%%%%%%%%%%%%%%%%%%%%%%%%%%%%%%%%%%%%%%%%%%%%%%%%%%%%%%%%%%

We now proceed with a more detailed examination of point A, by
performing a hysteresis loop. In figure 2 we show the cycle for the
space-like plaquette: $\bt^\prr$ is set to 1.0, while $\bt$ runs. 
The lattice volume used has been $4^5,$ the step in $\bt$ was 0.01 and 
200 sweeps have been made at each point before proceeding to the next one.
A big hysteresis loop appears between $\bt=0.56$ and $\bt=0.62,$ 
indicating a first order transition. Apart from the phase transition 
it is seen that the space-like plaquette is growing from a value close 
to zero to a value close to one. Let us just mention that the
transverse-like plaquette exhibits a quite similar hysteresis loop 
between the same values of $\bt$ and its value varies between 0.46 and 0.73.

%%%%%%%%%%%%%%%%%%%%%%%%%%%%%%%%%%%%%%%%%%%%%%%%%%%%%%%%%%%%%%%%
\begin{figure}
\centerline{\hbox{\psfig{figure=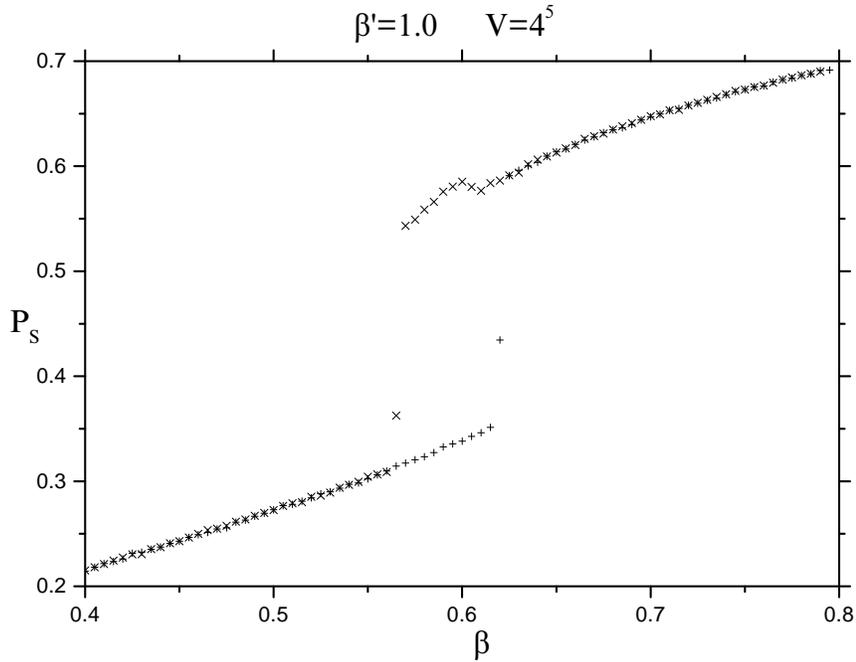,height=10cm}}}
\caption[f2]{Hysteresis loop for point A of figure 1. 
The quantity measured is the space-like plaquette.}
\label{f2}
\end{figure}
%%%%%%%%%%%%%%%%%%%%%%%%%%%%%%%%%%%%%%%%%%%%%%%%%%%%%%%%%%%%%%%%

Of course, more evidence is needed to substantiate our claim that the
phase transition from the strong to the Coulomb phase is of first order. 
This evidence is provided in figure 3, 
where we show a clear two-state signal in the distribution of the 
space-like plaquette. This is typical for the values of $\bt$ in the 
phase transition region.

%%%%%%%%%%%%%%%%%%%%%%%%%%%%%%%%%%%%%%%%%%%%%%%%%%%%%%%%%%%%%%%%
\begin{figure}
\centerline{\hbox{\psfig{figure=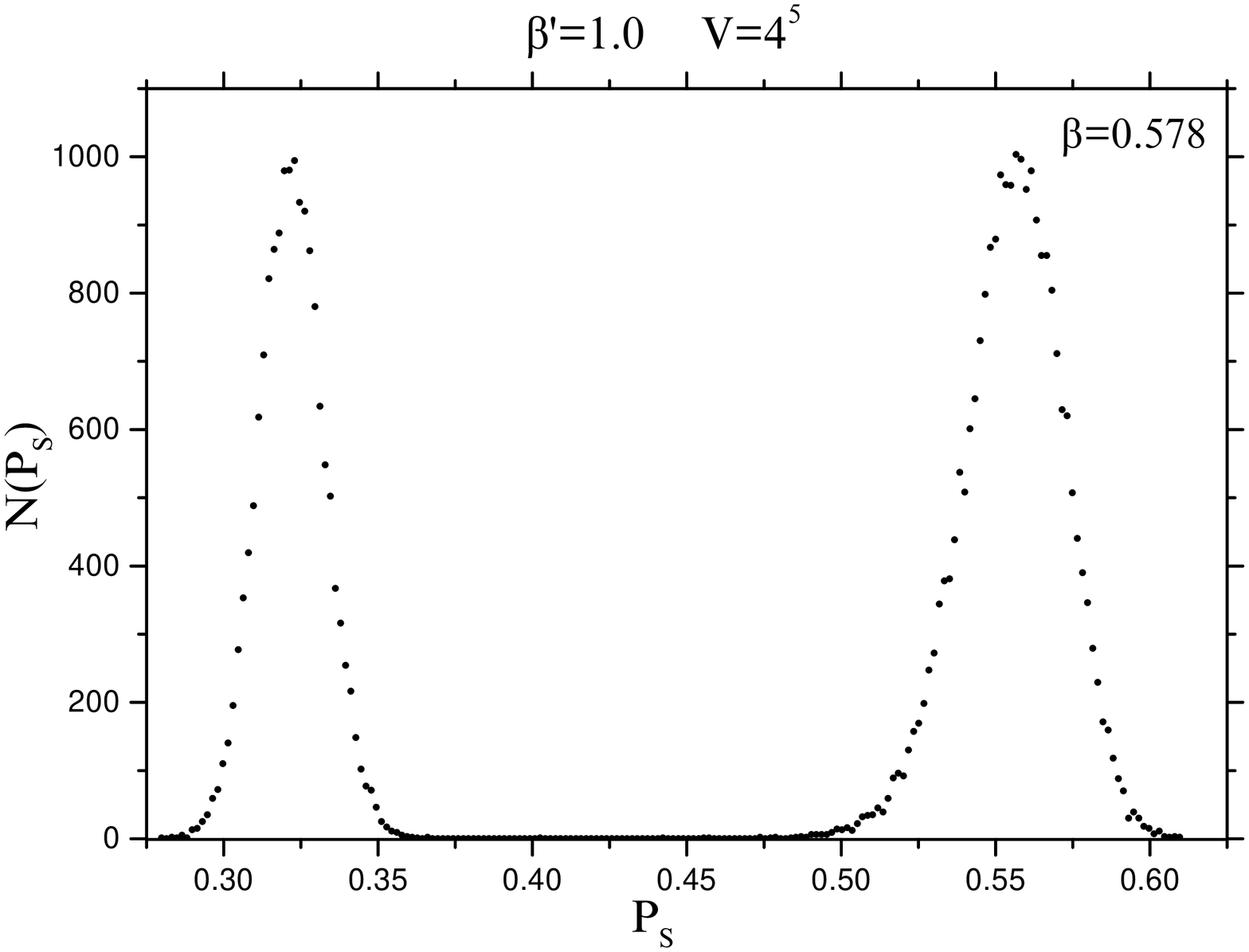,height=10cm}}}
\caption[f3]{A two-state signal for space-like plaquette for $\beta^{\prime}=1.0$ and  $V=4^{5}$ indicating a first order phase transition.} 
\label{f3}
\end{figure}
%%%%%%%%%%%%%%%%%%%%%%%%%%%%%%%%%%%%%%%%%%%%%%%%%%%%%%%%%%%%%%%%

The hysteresis loop relevant for point B of figure 1 is very weak, so we 
went on to a more detailed study. In figure 4 we show the mean values of the
space-like plaquette when $\bt^\prr$ is set to 0.2 and $\bt$ takes on
several values. The number of thermalization iterations used at each point 
varied between 10000 and 20000, while the measurements have required 30000 
to 50000 sweeps; only the results of one sweep out of five have been 
taken into account; this means that 6000-10000 measurements
have contributed to the mean values. 
Three volumes have been used: $4^5,~6^5,~8^5.$ The curve gets steeper as
the volume increases, so a reasonable guess would be that this
transition (from the strong to the layered phase) is second order.
We mention that the transverse-like plaquette has a small value (about 0.10)
in this parameter range. This confirms the picture that gauge fields
remain localized along the transverse direction, yielding a large 
corresponding string tension.

%%%%%%%%%%%%%%%%%%%%%%%%%%%%%%%%%%%%%%%%%%%%%%%%%%%%%%%%%%%%%%%%
\begin{figure}
\centerline{\hbox{\psfig{figure=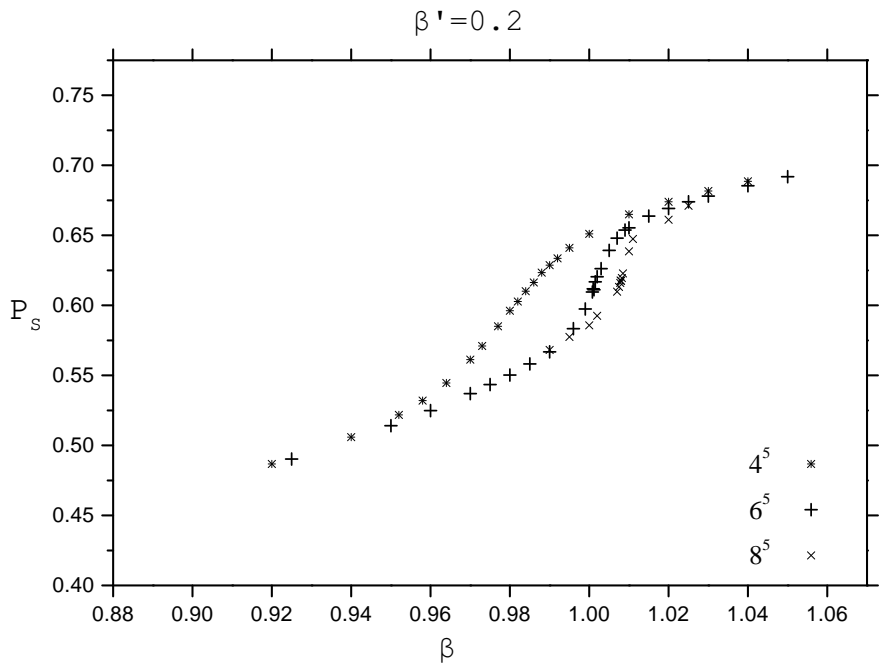,height=10cm}}}
\caption[f4]{Mean values for the space-like plaquette 
at point B of figure 1.}
\label{f4}
\end{figure}
%%%%%%%%%%%%%%%%%%%%%%%%%%%%%%%%%%%%%%%%%%%%%%%%%%%%%%%%%%%%%%%%

Figure 5 contains further elaboration on point B. We have computed the 
susceptibility of the space-like plaquette for the three lattice volumes 
$4^5,~6^5,~8^5$ to find the
volume dependence of the peak of the susceptibility. The number of sweeps 
needed are the same as the ones in figure 4.
It is well known that for a first order phase transition the peak
scales with the volume, while for a second order transition the
peak scales with a power of the volume smaller than one. The volume
ratios read: $\f{6^5}{4^5} \simeq 7.59,~\f{8^5}{4^5} =32.0.$
It is obvious that the ratios of the peaks $(\f{1.3}{0.6} \simeq 2.16,~
\f{1.75}{0.6} \simeq 2.92)$ are much smaller than the corresponding 
volume ratios, so the transition is of second (or higher) order for
the volumes that we have examined.
Thus we have here the very interesting possibility of the existence of
a continuum limit for this lattice theory along this phase transition line.
We comment here that in the $\bt^\prr \to 0$ limit we recover the 4-D U(1) 
pure gauge model. Recent simulations \cite{Jersak} (see also \cite{Jan}) 
have shown that the phase transition of the 4-D model is of second order, 
which matches very well with our result that the strong-layered phase 
transition is of second order. In particular, the 4-D U(1) phase transition 
may be considered as a shadow of the strong-layered phase transition.
We also note here that the phase transition point ($\bt \simeq 1.0$) 
is very close to the prediction of the one-loop mean field result 
(curve (II) of figure 1) and rather far from the zero order mean field 
(curve (I) of figure 1). This transition has also been characterized as second
order in \cite{arjan} by measuring hysteresis loops for 
Polyakov line correlators.

%%%%%%%%%%%%%%%%%%%%%%%%%%%%%%%%%%%%%%%%%%%%%%%%%%%%%%%%%%%%%%%%
\begin{figure}
\centerline{\hbox{\psfig{figure=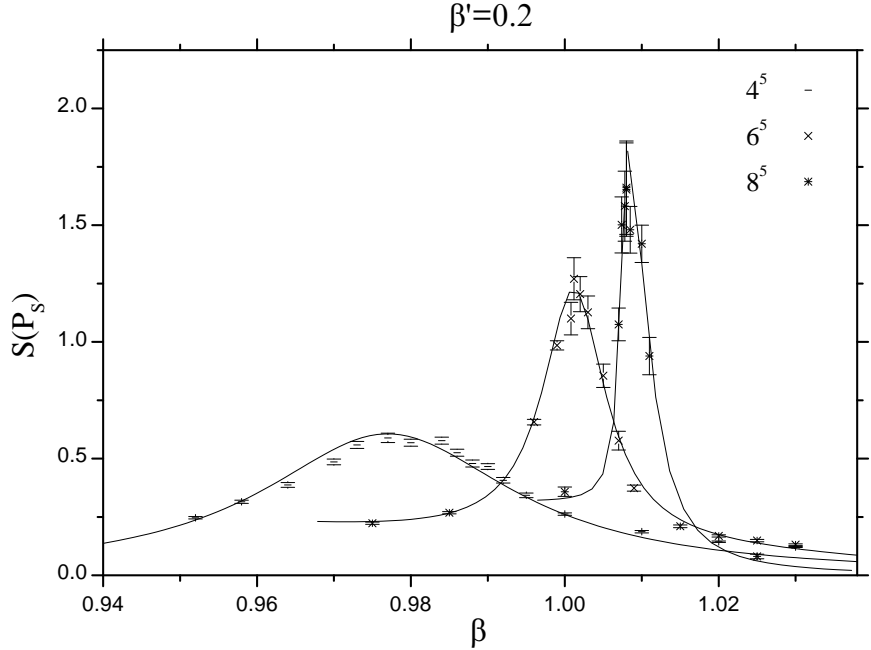,height=10cm}}}
\caption[f5]{Volume dependence of the susceptibility of the space-like
plaquette. The critical point B of figure 1 (strong-layered transition)
is relevant here.}
\label{f5}
\end{figure}
%%%%%%%%%%%%%%%%%%%%%%%%%%%%%%%%%%%%%%%%%%%%%%%%%%%%%%%%%%%%%%%%

In figure 6 we show the Binder cumulant of the space-like plaquette 
to have an additional check on the strength of the layered-Coulomb  
transition. For a first order transition the cumulant has a minimum at
the critical point, which retains its depth with increasing volume. 
A second order transition is characterized by the decrease of the 
depth of the dip. In our figure we observe that  the minimum get 
shallower with increasing volume, in agreement with the estimate 
based on the previous figure that the phase transition is of second order.

%%%%%%%%%%%%%%%%%%%%%%%%%%%%%%%%%%%%%%%%%%%%%%%%%%%%%%%%%%%%%%%%
\begin{figure}
\centerline{\hbox{\psfig{figure=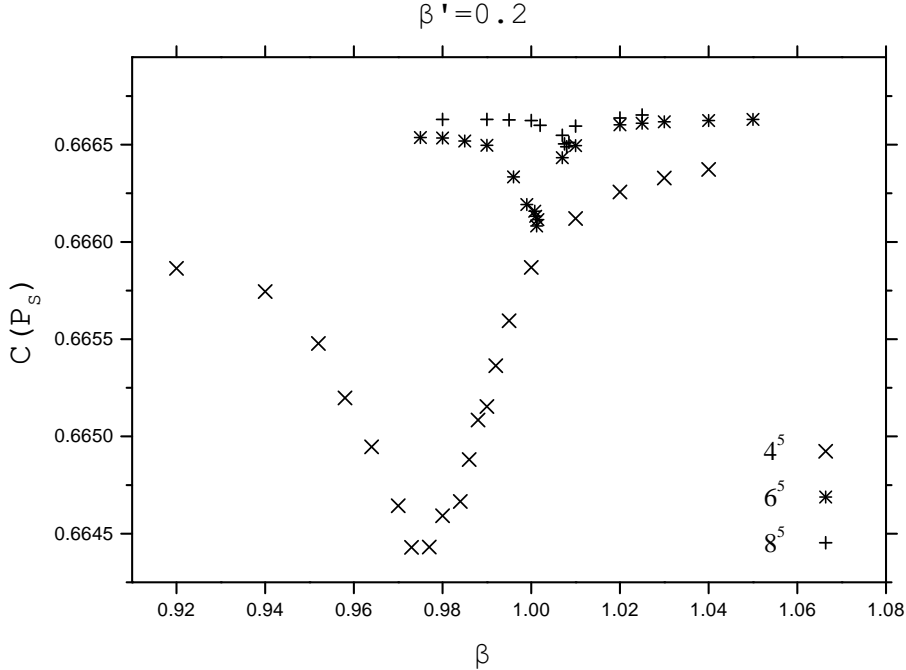,height=10cm}}}
\caption[f6]{Volume dependence of the Binder cumulant of the space-like
plaquette. Also here, as in the previous figure, the critical point B
of figure 1 (strong-layered transition) is relevant.}
\label{f6}
\end{figure}
%%%%%%%%%%%%%%%%%%%%%%%%%%%%%%%%%%%%%%%%%%%%%%%%%%%%%%%%%%%%%%%%

Figure 7 contains mean values in connection with point C of figure 1, 
which lies on the layered-Coulomb phase transition. 
The technical details about the number of sweeps are the same as the ones of 
figure 4 above, but the sensitive quantity is the transverse-like plaquette in 
this case. Also in this case we used three volumes. 
The curve grows slightly steeper
as the volume increases, but the change is very small. Another feature
is that the ``critical point" moves to smaller $\bt^\prr$ for increasing
volume, contrary to the tendency in figure 4. 
The space-like plaquette is almost constant, varying in the range 0.79-0.80;
this is due presumably to the fact that the change happening here is
a transition from 4-D to 5-D Coulomb phase, which is not expected to 
give a great change in the $1 \times 1$ plaquette behaviour.

%%%%%%%%%%%%%%%%%%%%%%%%%%%%%%%%%%%%%%%%%%%%%%%%%%%%%%%%%%%%%%%%
\begin{figure}
\centerline{\hbox{\psfig{figure=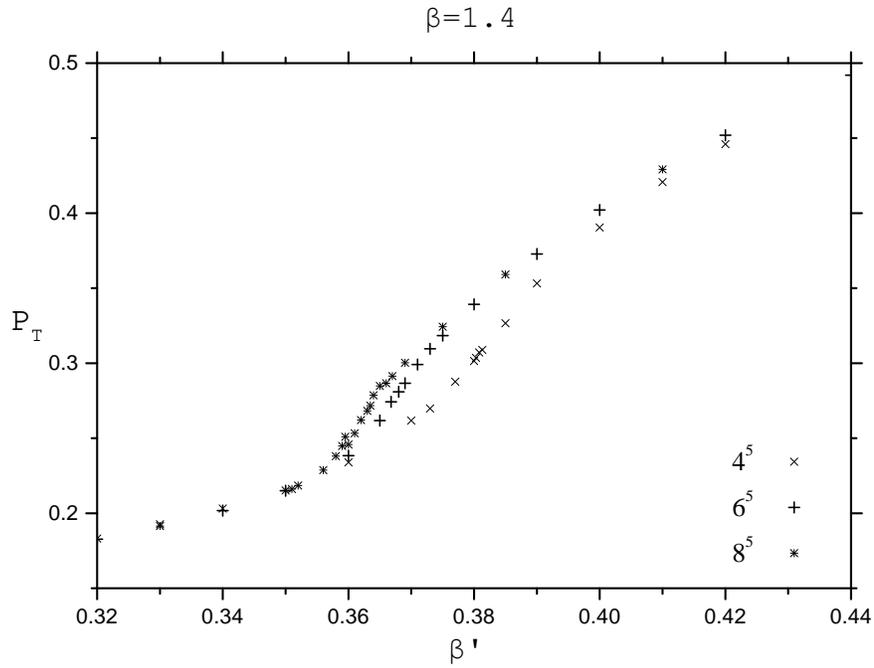,height=10cm}}}
\caption[f7]{Mean values of the transverse-like plaquette 
relevant for point C of figure 1 (layered-Coulomb 
transition).}
\label{f7}
\end{figure}
%%%%%%%%%%%%%%%%%%%%%%%%%%%%%%%%%%%%%%%%%%%%%%%%%%%%%%%%%%%%%%%%

Since the transition in the previous figure appears very weak, it 
deserves further more detailed study. In figure 8 we depict the volume 
dependence of the susceptibility of the transverse-like plaquette. The 
peaks of the susceptibilities seem to saturate to a final value of
about 1.8; in fact this value for the biggest volume is very close to the
value 1.4 for the smallest volume, although the volume ratio is 32!
It appears that the value of the peak saturates to a definite value,
independent of the volume; this could be a signal that the transition
is compatible with a crossover.

%%%%%%%%%%%%%%%%%%%%%%%%%%%%%%%%%%%%%%%%%%%%%%%%%%%%%%%%%%%%%%%%
\begin{figure}
\centerline{\hbox{\psfig{figure=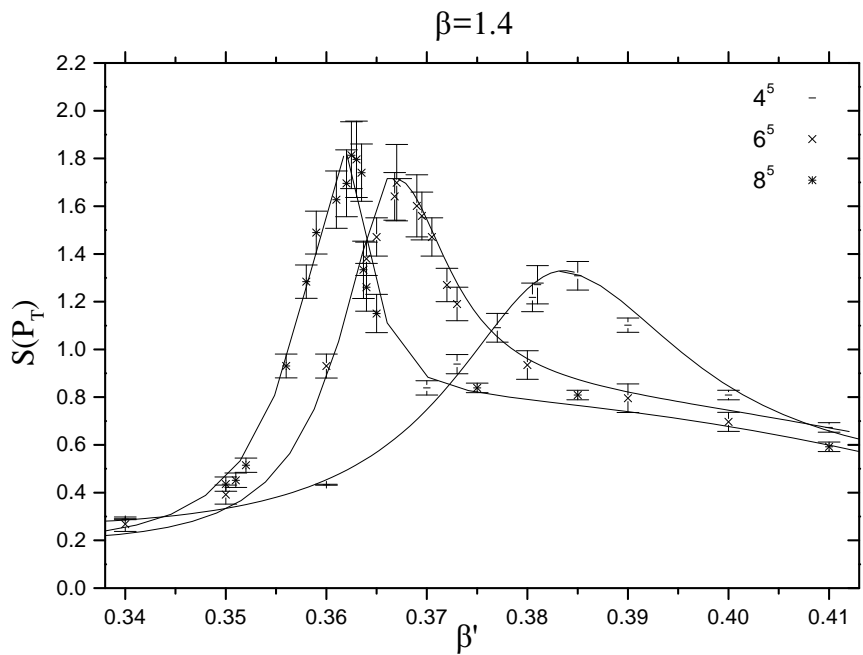,height=10cm}}}
\caption[f8]{Volume dependence of the susceptibility of the transverse-like
plaquette. The critical point C of figure 1 (layered-Coulomb transition)
is relevant here.}
\label{f8}
\end{figure}
%%%%%%%%%%%%%%%%%%%%%%%%%%%%%%%%%%%%%%%%%%%%%%%%%%%%%%%%%%%%%%%%

\subsection{Model II}

We now pass to model II, which is a model in which $\bt$ and $\bt^\prr$ are not
independent, as in model I, but are instead related as 
$$ \bt^\prr = \bt e^{-\lambda |n_T|}.$$
$\lambda$ is a positive parameter that we change by hand, while $|n_T|$ is the
(absolute) distance of the current 4-D subspace from a fixed 4-D subspace,
used as a reference layer, that is the origin of the ordinate along 
the transverse direction. Thus the coupling $\bt^\prr$ is not constant 
over the whole extent of the lattice, since it is meant to represent the 
Randall-Sundrum model.

We have already gained some experience with the behaviour of our model 
in the case of constant $\bt^\prr.$ We have seen there that the new 
``layered phase" makes its appearance at small values of the 
$\bt^\prr$ parameter. The physical meaning of the layered phase is that 
the space-like 4-D subspaces (named ``layers" for convenience) 
decouple and behave independently. Now, we expect the same phenomenon to
occur when $\bt^\prr$ takes small values not only over the whole lattice, 
but also {\em locally,} which is the case of model II. Suppose that we
start with the layer which has $n_T=0.$ Then $\bt^\prr = \bt.$ If we move 
to layers with $|n_T|>0,$ $\bt^\prr$ will decrease and it will get its 
minimum value when we reach the layer with the largest distance from the 
reference layer. Depending on the value of the $\lambda$ parameter, this
value of $\bt^\prr$ may be small enough that the layered phase shows up 
{\em locally.} The picture is that the layer with largest $|n_T|$ (the
most distant layer) will decouple, while its 
first neighbours (along the transverse direction) will still be connected 
with the rest of the lattice. If we increase $\lambda,$ we expect that 
not only the most distant layer, but also the next layer will have small
enough $\bt^\prr$ for decoupling. If the $\lambda$ parameter grows bigger and 
bigger we expect that more and more layers will decouple. 

We have used the mean values of the space-like and the transverse-like
plaquettes, but for model II their definitions are slightly different than
the previous ones. The reason is that, because of the 
dependence of $\bt^\prr$ on the transverse coordinate, we should consider 
these mean values for each layer separately. More precisely, the 
two operators read:
\be
{\rm Space-like~Plaquette:~~~} {\hat P}_s(|n_T|) \equiv \f{1}{12 N^4}
\sum_{x,1 \le \mu < \nu \le 4} \cos F_{\mu \nu}(x)|_{fixed~|n_T|}
\ee
\be
{\rm Transverse-like~Plaquette:~~~} {\hat P}_T (|n_T|) \equiv \f{1}{8 N^4}
\sum _{x,1 \le \mu \le 4} \cos F_{\mu T}(x)|_{fixed~|n_T|}
\ee
We have used the mean values $P_s(|n_T|)\equiv <{\hat P}_s>|_{fixed~|n_T|}$ 
and $P_t(x_t) \equiv <{\hat P}_t>|_{fixed~|n_T|}.$

We will study two cases of model II, corresponding to the cases of model I
that have already been examined. Figure 9 contains the space-like
plaquette ${\hat P}_s(|n_T|)$ 
versus $\lambda$ if $\bt^\prr =0.2$ (point B of figure 1) 
and the lattice volume is $6^5.$
Then $\bt = \bt^\prr e^{\lambda |n_T|}.$ 
For this particular lattice size, since the reference layer has $n_T=0,$ 
the two nearest neighbours have $|n_T|$ equal to $1$ and 
the next ones have $n_T$ equal to $2.$ 
We observe in the figure a decoupling transition at 
$\lambda$ about 0.8 for the first decoupling layer (the one with $|n_T|=2;$ 
upper curve). At this value of $\lambda$ the layers with $|n_T|=1$ 
have rather small values for $\bt,$ so they do not show any sign of
decoupling. However, when $\lambda$ grows even larger, we may see the
decoupling transition of these ``second layers" at $\lambda \simeq 1.65.$

%%%%%%%%%%%%%%%%%%%%%%%%%%%%%%%%%%%%%%%%%%%%%%%%%%%%%%%%%%%%%%%%
\begin{figure}
\centerline{\hbox{\psfig{figure=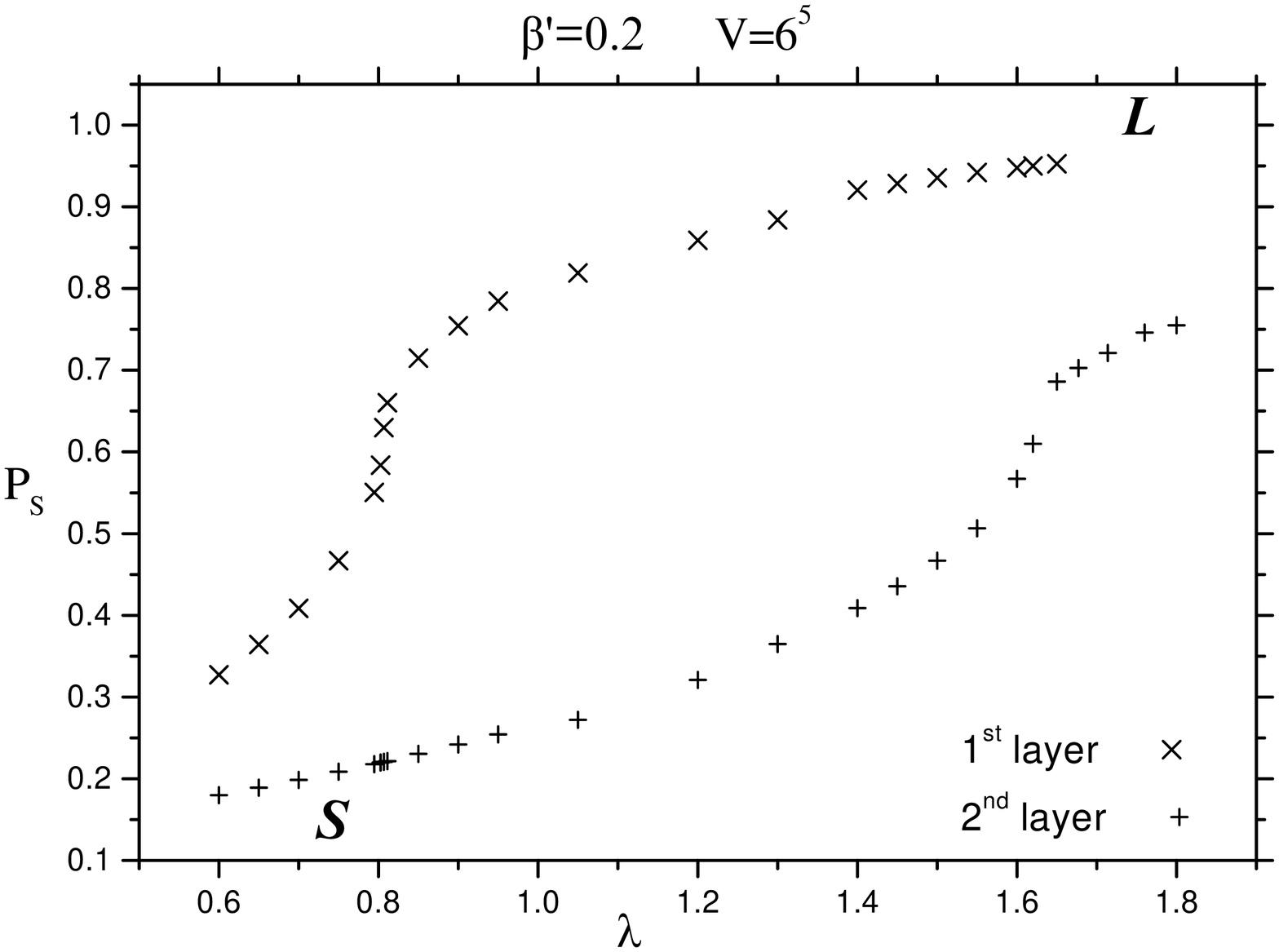,height=10cm}}}
\caption[f9]{Space-like plaquette versus $\lambda$ for the 
two values $|n_T|=2$ (denoted by ``first layer") and $|n_T|=1$ 
(denoted by ``second layer"). $\bt^\prr$ is set to 0.2, so we are 
probing the strong-layered transition. The layered phase lies 
in the right part of the graph and the strong phase in the left 
part of it. }
\label{f9}
\end{figure}
%%%%%%%%%%%%%%%%%%%%%%%%%%%%%%%%%%%%%%%%%%%%%%%%%%%%%%%%%%%%%%%%

A very interesting question is the relation of the layer decoupling 
phase transitions observed in figure 9 to the corresponding phase 
transitions of model I. To this end we use the relation between 
$\bt,~\bt^\prr$ and $\lambda$ and construct figure 10; 
this figure is similar to the previous one, the difference being that the 
horizontal axis is not $\lambda,$ but rather the related variable $\bt.$
In addition, we have put on this graph the results for the mean values
of the space-like plaquette for model I (reproduced from figure 3) and
the results coming from processing the data of figure 9. The lattice
volume $6^5$ has been used for all data in figure 10. A striking effect takes
place: all three curves lie upon one another. Thus, presumably, the 
phase transition of model II is nothing more than the corresponding phase 
transition of model I, if both are expressed in terms of $\bt$
and $\bt^\prr.$ In other words, if we revisit the previous figure 9,
from the position of $\lambda$ for the first decoupling transition 
one may determine the values of $\lambda$ for the remaining decoupling
transitions, since they correspond to the same value of $\bt.$
It is plausible that the decoupling phase transitions are just the 
``local" version of the phase transitions which appear in model I: there 
the parameter $\bt^\prr$ has the same value throughout the lattice, so all
layers were decoupled from each other simultaneously when $\bt$ 
took a suitable value. In model II a similar phenomenon takes place
{\em locally:} the layers do not decouple simultaneously but one after the
other.

%%%%%%%%%%%%%%%%%%%%%%%%%%%%%%%%%%%%%%%%%%%%%%%%%%%%%%%%%%%%%%%%
\begin{figure}
\centerline{\hbox{\psfig{figure=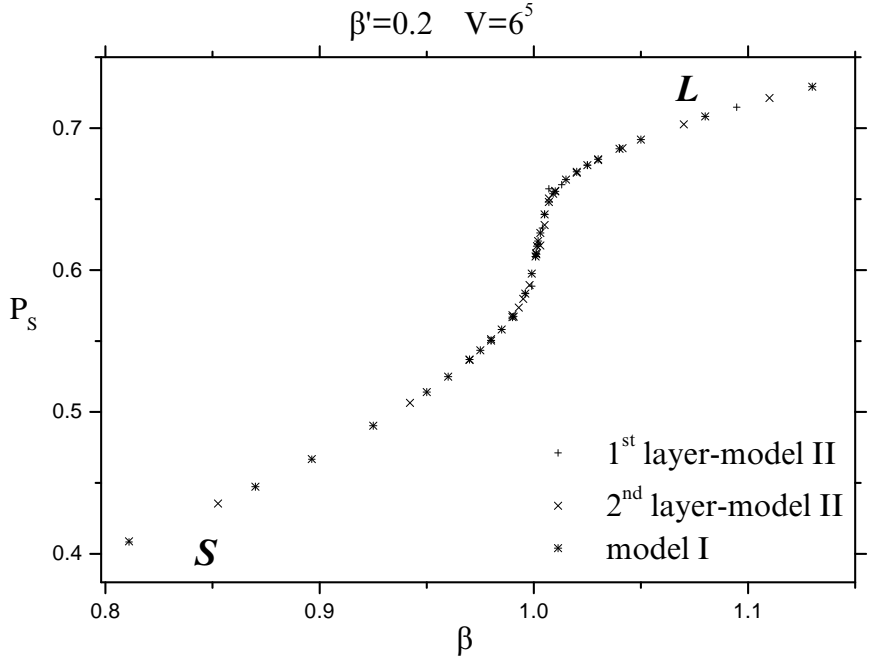,height=10cm}}}
\caption[f10]{The data of figure 9 plotted versus 
$\beta.$ }
\label{f10}
\end{figure}
%%%%%%%%%%%%%%%%%%%%%%%%%%%%%%%%%%%%%%%%%%%%%%%%%%%%%%%%%%%%%%%%

Figure 11 is analogous to figure 9, but corresponds to the 
layered-Coulomb transition (point C of figure 1) rather than the 
strong-layered one. As a consequence, the relevant quantity is the
transverse-like plaquette $P_T(|n_T|).$ The parameter $\bt$ is set to 1.4 and 
$\bt^\prr$ is found through the relation: $\bt^\prr = \bt e^{-\lambda |n_T|}.$
This means that $\bt^\prr$ at each layer decrease with $\lambda,$ so
the corresponding transverse-like plaquette should decrease, 
which is what we observe. Thus the left part of the graph corresponds 
roughly to the Coulomb phase, while its right part to the
layered phase of figure 1. We observe similar phenomena to figure 9, 
that is the most distant layer decouples at $\lambda \approx 0.7,$ 
and the next at $\lambda \approx 1.3.$

%%%%%%%%%%%%%%%%%%%%%%%%%%%%%%%%%%%%%%%%%%%%%%%%%%%%%%%%%%%%%%%%
\begin{figure}
\centerline{\hbox{\psfig{figure=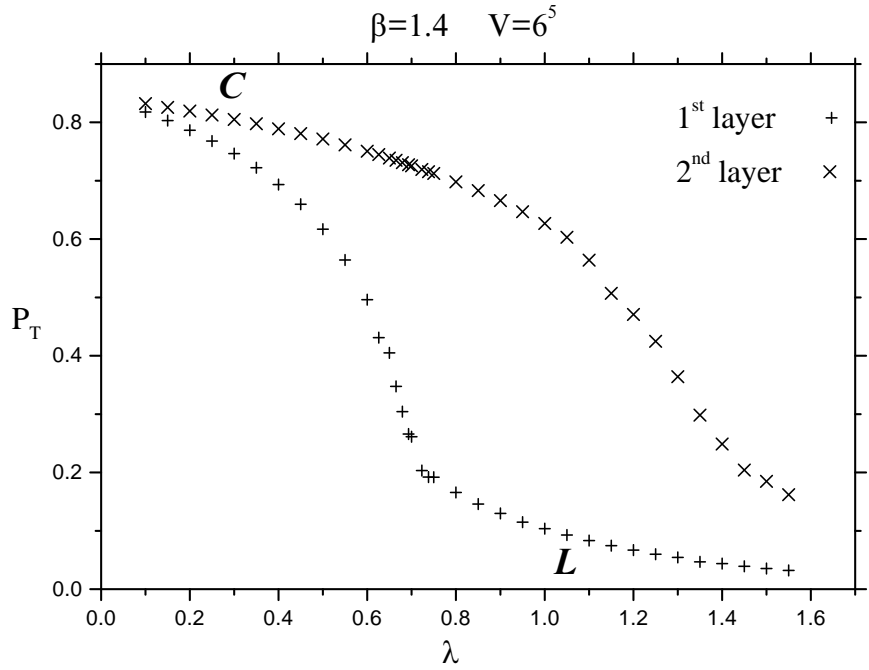,height=10cm}}}
\caption[f11]{Space-like plaquette versus $\lambda$ for the 
two values $|n_T|=2$ (denoted by ``first layer") and $|n_T|=1$
(denoted by ``second layer"). $\bt$ is set to 1.4, so we are 
probing the layered-Coulomb transition. The layered phase lies
in the right part of the graph and the Coulomb phase in the left 
part of it.}
\label{f11}
\end{figure}
%%%%%%%%%%%%%%%%%%%%%%%%%%%%%%%%%%%%%%%%%%%%%%%%%%%%%%%%%%%%%%%%

Figure 12 is a redrawing of figure 11 in the same way as figure 10 was 
the redrawing of figure 9. The only difference between figures 11 and 12 
is that the mean values are plotted versus $\bt^\prr$ rather than
versus $\lambda.$ It shows once more that the phase transitions take place 
at the same values of $\bt^\prr;$ moreover this value coincides with the
critical point of model I. Thus, once more, we find no new phase transitions,
but only a ``local" version of the old ones; so there is no reason to 
study this phase transition in more detail, since presumably its 
characteristics will be the same as the ones of the ``global" transition.

%%%%%%%%%%%%%%%%%%%%%%%%%%%%%%%%%%%%%%%%%%%%%%%%%%%%%%%%%%%%%%%%
\begin{figure}
\centerline{\hbox{\psfig{figure=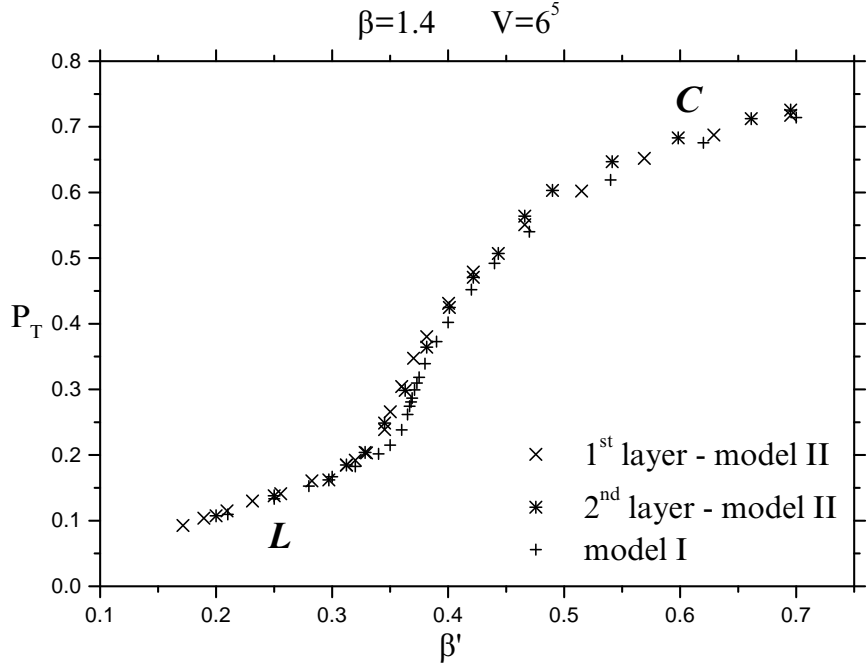,height=10cm}}}
\caption[f12]{The data of figure 11 plotted versus 
$\beta.$ }
\label{f12}
\end{figure}
%%%%%%%%%%%%%%%%%%%%%%%%%%%%%%%%%%%%%%%%%%%%%%%%%%%%%%%%%%%%%%%%

\section{Conclusions}

 We have studied here the problem of gauge field localization in the case of 
anisotropic gauge couplings. We have considered two models, model I and model 
II. The former has couplings independent from each other and 
the coordinates, while the latter has an exponentially bigger coupling
in the fifth direction. The model I may be realized in a continuum flat 
homogeneous and anisotropic space, while the model II may be realized in 
a curved $AdS$ space. Although scalars, fermions as well as  gravitons may 
be localized on a three-brane, at least in the RS background, there is an 
open question for the gauge bosons. The reason
  is that a massless gauge field cannot be localized on a brane since 
it does not depend on the extra dimensions.  One of course may 
imagine a situation in which a Higgs mechanism makes the gauge bosons massive
 in the bulk and massless inside the brane. However, in this case,
a kind of ``no-go theorem" does not allow for massless photons on the brane,
due to the
Meissner effect. 
%the bulk is superconductive and 
%the gauge theory on the brane is not in the Coulomb phase.
 In fact,
 electric fields along the brane dies off exponentially so that there is no
Coulomb phase on the brane. A way out is   
to consider a dual picture where the bulk is in a confining phase 
(``confining medium" with monopole condensation). In this case
the gauge theory on the brane is in a Coulomb phase with massless photons. 

In our case, we have considered a five-dimensional pure $U(1)$ 
gauge theory in the strong phase. 
We have shown that there is a second order phase transition from the strong 
phase to a layer phase. The gauge theory along the layer is in the Coulomb 
phase, while it is still in the strong in the transverse direction. Thus, a 
four-dimensional massless photon exists on the four-dimensonal layers 
indicating
that localization of the bulk gauge fields has been achieved. 
This is consistent 
with the  no-go theorem as the gauge theory in the bulk is  not in a 
Higgs phase. Contrary, it is in a strong phase and in a sense 
may be viewed as 
realization of the ``confining medium" proposal. 
It should be stressed that one 
cannot see these effects simply by solving bulk Maxwell equations since the 
localization of the gauge field is a non-perturbative effect in which 
strong dynamics is involved. Clearly, such dynamics cannot be captured by
any perturbative analysis.

Another  mechanism for gauge field localization would be, for
example, an  $SU(2)$-Higgs 
model where the theory is in the strong symmetric phase 
 in the bulk and in the Higgs phase on the  brane. Moreover, such a picture 
may also be consistent with a supersymmetric theory in the bulk and broken 
supersymmetry on the brane.  

However, we should keep in mind that although we know that there exists a
second order phase transition between the strong and layer phases of the 
five-dimensional gauge theory, we have not made any detailed study 
of the continuum limit yet. 
 
\vspace{1cm}

{\Large{\bf Acknowledgements}}

\vspace{0.5cm}

This work is partially supported by the TMR projects ``Finite temperature
phase transitions in Particle Physics", EU contact number: FMRX-CT97-0122,
RTN contract RTN-99-0160 and the $\Gamma.\Gamma.E.T$ grand $97E\L/71$. 
Stimulating discussions with F. Karsch, C.P. Korthals-Altes, 
S. Nicolis and N. Tetradis are gratefully acknowledged. We also thank  
V. Stergiou and N. D.Tracas for helpful advice on plotting the numerical data.

\end{document}